# Conservation of transverse orbital angular momentum for spatiotemporal optical vortices

Jordan Adams, Youngbin Park[1, 2, 3], and Andy Chong[1, 2, 3*]

[1]Department of Physics, Pusan National University, Busan, 46241, Republic of Korea
[2]Institute for Future Earth, Pusan National University, Busan, 46241, Republic of Korea
[3]Quantum Science Technology Center, Pusan National University, Busan, 46241, Republic of Korea
*chong0422@pusan.ac.kr

## Abstract

The transverse orbital angular momentum (OAM) of spatiotemporal optical vortices (STOVs) has been a topic of active debate in recent years. Previous studies, relying on narrowband and paraxial approximations, resulted in unprecedented conclusions. Some researchers have proposed that electromagnetic waves require a modified definition of OAM to ensure conservation. In this work, we demonstrate that the transverse OAM of STOVs is highly sensitive to these commonly adopted approximations. Using the standard definition of OAM for electromagnetic waves, we demonstrate that the total transverse OAM is conserved during propagation once the narrowband and paraxial assumptions are removed.

## 1 Introduction

Light waves that possess orbital angular momentum (OAM) have been intensively researched for decades [1,2]. These optical beams have an azimuthally increasing phase across the beam profile which results in a phase singularity and intensity null at the center. Optical vortices (OVs) have proven to be beneficial for various applications such as free-space communications [3]-[5], particle manipulation [6]-[8], and even additive manufacturing [9],[10].

Spatiotemporal optical vortices (STOVs) are a newer type of vortices with growing interest [11], [12]. In contrast to OVs, STOVs are oriented perpendicular to the propagation direction and possess transverse OAM. STOVs were first experimentally observed in nonlinear pulse propagation [13], and subsequent studies showed that they could be readily generated using pulse shaping techniques [14], [15].

Despite years of extensive research, the value and proper method for calculating the transverse OAM of STOVs remain subjects of ongoing debate. In free-space propagation, STOVs appear to diffract and expand only along the transverse spatial dimensions, while the time profile and local spatial profile parallel to the propagation direction remain largely unaffected. As a result, STOVs undergoing free-space propagation are distorted into multi-lobed structures resembling Hermite-Gaussian profiles or vice-versa.

Consistent with this distortion, the OAM value appears to change during propagation, which seems to contradict the principle of OAM conservation. It was proposed that the correct method to calculate the OAM of a STOV is to exclude any momentum contributions along the propagation axis $z$ or local time $\zeta = v_g t - z$, where $v_g$ is the group velocity. This approach yields unique half-integer topological charge $\frac{l}{2}$ OAM values, where $l$ is the topological charge of the spatiotemporal azimuthal phase of $e^{il\theta}$ [16]-[20]. Removing the linear momentum ($P_z$ or $P_\zeta$) component along the propagation in the OAM calculation ensures the OAM conservation as the STOV propagates paraxially [18]. Although this approach guarantees the OAM conservation during the paraxial propagation, it introduces an unprecedented method of calculating OAM. This method implies that the OAM of light should be treated distinctly from the conventional OAM calculation.

Others have proposed using a traditional OAM calculation including $P_\zeta$, combined with shifting the energy center to recover a whole integer $l$ OAM values [21], [22]. However, even with this adjustment, the OAM still changes with paraxial propagation. Another approach concluded that the net transverse OAM is zero, proposing that intrinsic OAM is canceled by extrinsic OAM due to the STOV propagation [23], [24]. However, other studies have shown that transverse torque can be exchanged between light and matter via STOVs, challenging the validity of this approach [18], [19].

Here, we show that the OAM calculation is highly sensitive to commonly used plausible approximations. Previous results have relied on paraxial equations with a small temporal-frequency bandwidth assumption to study STOVs analytically. While these approximations could provide elegant analytical calculations, we find that the OAM calculation is highly sensitive since these approximations often fail to give accurate results. In our analysis, removing these approximations reveals that the OAM remains constant, reaffirming its conservation of OAM during propagation.

The appendix of Ref [18] calculated the OAM of a STOV using numerical non-paraxial propagation simulation and observed that the OAM changes during propagation. In contrast, we show the OAM calculation is also highly sensitive to the resolution of the numerical simulation. Apparent changes in OAM decreases as the numerical simulation resolution increases. Therefore, to achieve accurate results require appropriate approximations under suitable simulation resolutions.

To understand the conservation of transverse OAM in STOVs, we first examine the astigmatic focusing of an OV with commonly used approximations, where the beam profile rapidly changes along one axis while remaining nearly constant along the other. We then present the STOV OAM evolution during the free-space propagation with different approximation conditions. Since the STOV diffracts primarily along one spatial direction, its propagation behavior is analogous to that of an astigmatically focused OV. Analyzing the STOV evolution during propagation, with and without approximation, shows that the distortions due to the frequency dependent diffraction occurs along the propagation direction. Counting such effects by carefully removing some common approximations ensure the conservation of the transverse OAM in the same manner as an astigmatically focused OV.

We believe this work will deepen the understanding of transverse OAM in light by affirming its conservation within the framework of the conventional definition of OAM. We also believe this work will be an important guide for the transverse OAM-based light-matter interactions and free-space communication, which rely on accurate calculations of OAM.

## 2 Results

### 2.1 Optical vortex cylindrical focusing

To understand how OAM is conserved for a STOV, we begin by examining the OAM of an OV under astigmatic focusing. When an OV is astigmatically focused, for example by a cylindrical lens, a Hermite-Gaussian (HG) like profile forms at the focus. Likewise, when a HG beam is astigmatically focused, an OV appears at the focus. For this analysis, we assumed a specific beam profile at the focal plane, located at z=0 in terms of spatial-frequency components as in Equation (1). The varying beam sizes along the x-and y-directions are assumed to simulate the effect of astigmatic focusing. Since the beam starts from the focal plane, the free-space propagates of the profile given in Equation (1) is analogous to the astigmatically defocusing case. However, the focusing case can be readily interpreted as the back-propagation of the profile give in Equation (1).

$$E(k_x, k_y, z = 0) = \left(k_x + k_y(A + iB)\right) \exp\left(-\left(\frac{k_x}{w_{k_x}}\right)^2 - \left(\frac{k_y}{w_{k_y}}\right)^2\right) \qquad (1)$$

A and B are unitless coefficients. When $A = 1, B = 0$ the field at the focus has a HG profile. The far-field evolution of this HG profile after the focus, corresponding to astigmatic defocusing, results in an OV. Conversely, when $A = 0, B = 1$, an OV is formed at the focus, and an HG profile will be formed at far-field under astigmatic defocusing. To determine the field evolution in propagation, we employ three approaches. The first approach is the angular spectrum method (ASM) which is an exact solution to the wave-equation [25]. The ASM is described by the following equation

$$E(k_x, k_y, z) = E_o \exp\left(-i2\pi z \sqrt{\left(\frac{\omega_o}{c}\right)^2 - k_x^2 - k_y^2}\right) \qquad (2)$$

where $E_o = E(k_x, k_y, z = 0)$ is the initial field at the focus. The spatial field is simply the inverse Fourier transform, $E(x, y, z) = \mathcal{F}^{-1}\left(E(k_x, k_y, z)\right)$. The second approach is the propagation with a paraxial approximation, which assumes that the beam is mostly collimated with small $k_x$ and $k_y$ components.

$$E(k_x, k_y, \omega, z) = E_o \exp\left(-i2\pi z \left(\frac{\omega_o}{c} - \frac{c}{2\omega_o}(k_x^2 + k_y^2)\right)\right) \qquad (3)$$

Under the paraxial approximation, the field evolution is described by Equation (3), which is obtained via a Taylor expansion of the square root term in Equation (2). Equation (3) is widely used to predict the propagation behavior when the beam size is much larger than the wavelength. Even though there is no further approximation is necessary to predict the propagation behavior, one can introduce an additional approximation. In case of an extreme astigmatic focusing, it is reasonable to leave out the diffraction in one direction, as the beam expands much slower. This third approximation gives

$$E(k_x, k_y, z) = E_o \exp\left(-i2\pi z \left(\frac{\omega_o}{c} - \frac{c}{2\omega_o}k_x^2\right)\right). \qquad (4)$$

For Equation (2), a closed-form solution of $E(x, y, z)$ is not available. However, $E(x, y, z)$ can be found analytically for both approximations (Equation (3) and (4)) allowing analytic expressions. The total energy-normalized OAM flux oriented along the propagation longitudinal direction is given by:

$$L_z = \frac{\iiint (xP_y - yP_x)\, dxdy}{\iiint |E|^2 dxdy} \qquad (5)$$

where $P_x = imag(E^* \frac{d}{dx} E)$ and $P_y = imag(E^* \frac{d}{dy} E)$ are the linear momentum in the direction of $x$ and $y$ respectively. Equation (5) provides a dimensionless normalized OAM instead of the absolute OAM. However, this dimensionless OAM is sufficient to track how the OAM changes. For the case of ignoring diffraction along y (Equation (4)), the longitudinal OAM is given by:

$$L_z = \frac{B\left(w_{k_x}^2 + w_{k_y}^2\right) - \frac{\pi A w_0 w_{k_y}^2 w_{k_x}^2 z}{c}}{w_{k_y}^2(A^2 + B^2) + w_{k_x}^2} \tag{6}$$

This shows the OAM changes linearly with propagation whenever $A$ is not zero and the field at the focus contains partial HG component. In other words, OAM is conserved under Equation (4) only when the field is a pure OV. However, with including diffraction in $k_y$ (Equation (3)) the OAM is

$$L_z = \frac{B\left(w_{k_y}^2 + w_{k_x}^2\right)}{w_{k_y}^2(A^2 + B^2) + w_{k_x}^2} \tag{7}$$

which is independent of $z$ for all field configurations. Results are shown in Figure 1 with numerical simulations using the ASM without approximations. The beam width of x-direction is set much smaller to simulate the focusing only along x-direction. Although the beam width in y-direction appears nearly static, regardless of whether the y-direction diffraction is included or not (Figure 1 (c) and (d)), there is a significant difference in the longitudinal OAM. Ignoring diffraction in the y-direction leads in a linear change in OAM. This indicates that the seemingly negligible expansion of the beam profile and the subtle wavefront evolution along $y$-direction are critical to ensure OAM conservation.

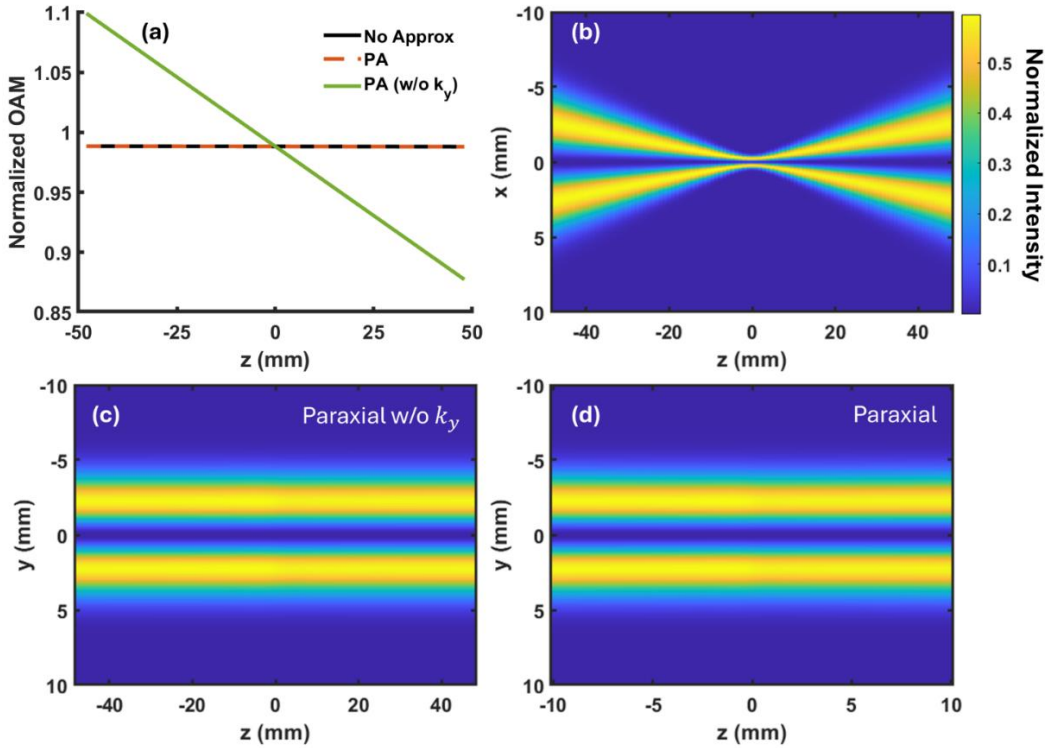

**Figure 1:** Optical vortex under astigmatic focusing for $A = B = 1$, $w_{k_y} = 1$ mm$^{-1}$, $w_{k_x} = 0.1$ mm$^{-1}$. (a) Ignoring diffraction along the $y$ direction results in a linear change in total OAM (labeled Paraxial Approximation (PA) w/o $k_y$). Including diffraction along both dimensions results in OAM that does not change with propagation with both paraxial approximation and without approximation. (b) $y = 0$ mm cross-section showing focusing in $x$. (c) $x = 0$ mm cross-section for paraxial propagation without including $y$ diffraction (eq. 4) and (d) with including $y$ diffraction (eq. 3) which shows no focusing in $y$ occurs and there are no apparent differences between the two cases.

## 2.2 Spatiotemporal optical vortex propagation

The previous section showed that neglecting diffraction and beam expansion in one-dimension can make the OAM appears to change during propagation. We now demonstrate the total transverse OAM of a STOV may appear to change during propagation, as a consequence of employing the paraxial and small-bandwidth approximations, even though it is in fact conserved. We consider the $k$-$\omega$ space representation of a STOV at a focus ($z = 0$) which has the general form of

$$E(k_x, k_y, \omega, z = 0) = \left(k_x + (\omega - \omega_o)(A + Bi)\right) \exp\left(-\frac{(k_x^2 + k_y^2)}{w_k^2} - \frac{(\omega - \omega_o)^2}{w_\omega^2}\right) \quad (8)$$

where $w_k$ is the spatial-frequency bandwidth in both the $x$- and $y$-directions and $w_\omega$ represents the frequency bandwidth. Symmetric diffraction is assumed for both transverse directions. Since free-space propagation does not introduce dispersion, astigmatic propagation naturally occurs. Here, we use the unit of mm$^{-1}$THz$^{-1}$ for A and B to balance the $k_x$ and $\omega - \omega_o$ terms. Similar to the OV case, when $A = 1$ mm$^{-1}$THz$^{-1}$, $B = 0$, the field at the focus exhibit a HG-like spatiotemporal profile, which evolves into a STOV in the far field. Conversely, for $A = 0, B = 1$ mm$^{-1}$THz$^{-1}$ the field is a STOV at the focus and evolves into the spatiotemporal HG in the far-field.

As in the previous section, we present results for using three approaches. First, the ASM without approximation:

$$E(k_x, k_y, \omega, z) = E_o \exp\left(-i2\pi z \sqrt{\left(\frac{\omega}{c}\right)^2 - k_x^2 - k_y^2}\right) \quad (9)$$

where $E_o = E(k_x, k_y, \omega, z = 0)$ is the initial field at the focus. This is identical to Equation (2), except $\omega_o$ has been replaced with $\omega$ to account for the polychromatic field. The second approach is to use the paraxial approximation which assumes $k_x$ and $k_y$ are small values:

$$E(k_x, k_y, \omega, z) = E_o \exp\left(-i2\pi z \left(\frac{\omega}{c} - \frac{c}{2\omega}(k_x^2 + k_y^2)\right)\right) \quad (10)$$

The final approach is the small bandwidth approximation which further assumes the center frequency $\omega_o$ is far greater than the temporal frequency bandwidth $w_\omega$. This allows us to set $\omega \approx \omega_o$ on the second term of Equation (10):

$$E(k_x, k_y, \omega, z) = E_o \exp\left(-i2\pi z \left(\frac{\omega}{c} - \frac{c}{2\omega_o}(k_x^2 + k_y^2)\right)\right). \quad (11)$$

Of course, the spatiotemporally wave packet profiles are obtained by the inverse Fourier transform of Equations (9)-(11). Once the spatiotemporally profiles are obtained, one can calculate the energy-normalized transverse OAM for each method using the formula:

$$L_y = \frac{\iiint (zP_x - xP_z)\, dxdydz}{\iiint |E|^2 dxdydz} \quad (12)$$

where $P_x = imag(E^* \frac{d}{dx} E)$ and $P_z = imag(E^* \frac{d}{dz} E)$ are linear momentum along $x$ and $z$. Since the Equation (12) is developed for a monochromatic optical wave, it is not strictly applicable to calculate OAM for polychromatic waves. However, we still employ Equation (12) in this manuscript for several reasons. First, it effectively captures the variations in OAM during propagation under each approximation. Second, existing studies on evaluating OAM in polychromatic waves are scarce. The resulting OAM can only be found analytically by applying both paraxial and small bandwidth approximations using Equation (11). Under these conditions, the OAM for propagation is given by

$$L_y = \frac{B\,c^4\,w_{k_x}^4 - 2\,A\,t\,\pi\,c^2\,\omega_0\,w_\omega^2\,w_{k_x}^2 + 2\,B\,\omega_0^2\,w_\omega^2}{2\,c\,\omega_0^2\,\left((A^2 + B^2)\,w_\omega^2 + w_{k_x}^2\right)} \tag{13}$$

where $t$ denotes absolute time, representing the propagation in the equation.

Similar to section 2.1, when $A \neq 0$, there is a linear change in OAM occurs over time during the propagation when both paraxial and small bandwidth approximations are used. This indicates that the transverse OAM of a collimated STOV is not conserved when propagating using paraxial low-bandwidth approximations.

Now, we show that propagating with ASM (Equation (9)) preserves the OAM all configurations of the field described by Equation (8). For a STOV at the focus, with parameters of $A = 0$ mm$^{-1}$THz$^{-1}$, $B = 1$ mm$^{-1}$THz$^{-1}$ and plugging in $w_k = 1/$mm, $w_\omega = 1$ THz, and $c = 0.3$ mm/ps gives the normalized OAM as $L_y \approx \frac{1}{2c} [\frac{mm}{ps}] = 1.667$. Figure 2 (a) shows iso-intensity plots of propagating the HG-like spatiotemporal profile propagated using Equation (9), illustrating the formation of the STOV. Figure 2 (b) shows that OAM is conserved for this case of the ideal STOV at the focus ($A = 0$, $B = 1$ mm$^{-1}$THz$^{-1}$), regardless of the method of propagation. In contrast, for a case of $A = 1$ mm$^{-1}$THz$^{-1}$, $B = 1$ mm$^{-1}$THz$^{-1}$, the OAM changes linearly with time when propagating using the paraxial and small-bandwidth approximations (Figure 2 (c)). Even when removing the small-bandwidth approximation, there is still a small linear change in OAM over time as is shown in Figure 2 (d). Propagating using ASM (Equation (9)) without any approximations yields a constant OAM in propagation.

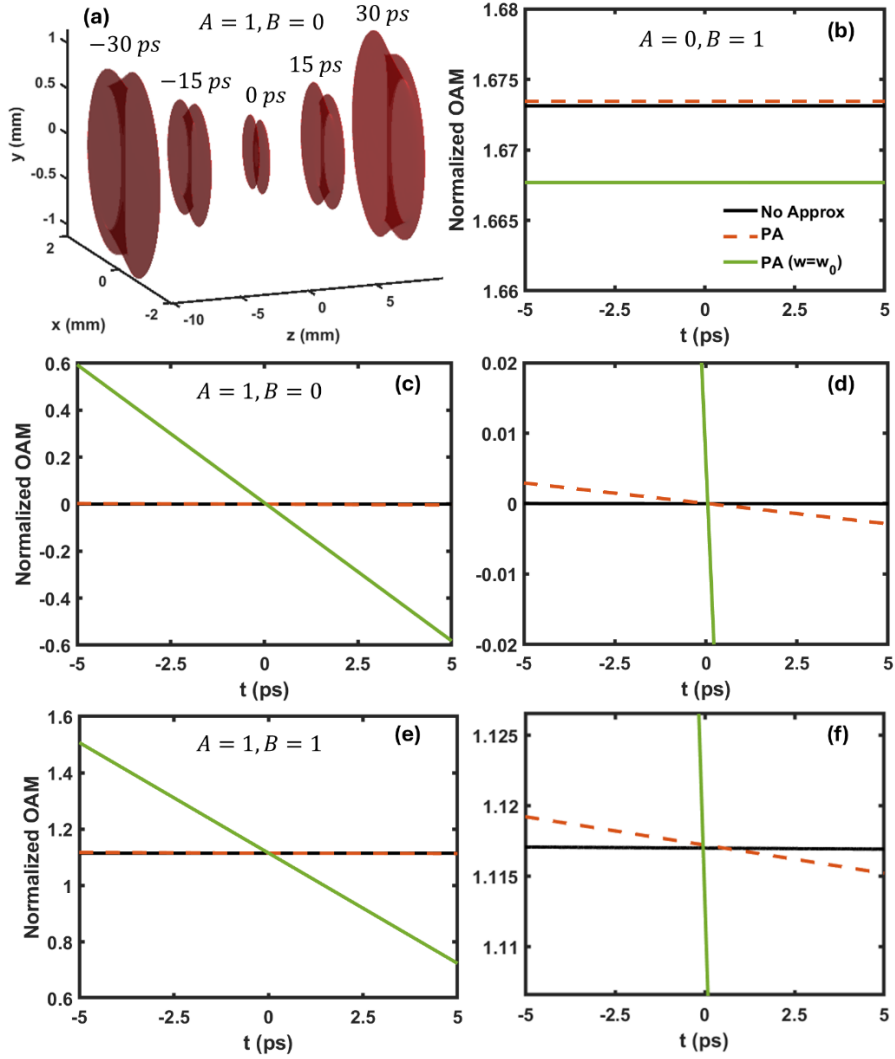

**Figure 2:** STOV transverse OAM conservation. **(a)** Iso-intensity profiles of a STOV ($A = 1mm^{-1}THz^{-1}, B = 0$) propagating through a focus as time advances. **(b)** OAM for each propagation approach for ($A = 0, B = 1mm^{-1}THz^{-1}$). **(c)** OAM for approach for ($A = 1mm^{-1}THz^{-1}, B = 0$). **(d)** The same results, shown over a smaller range of OAM values to show that only the approach with no approximation results in full OAM conservation. **(e)** OAM for approach for ($A = 1mm^{-1}THz^{-1}, B = 1mm^{-1}THz^{-1}$). **(f)** The same results, shown over a smaller range of OAM values to show that only the approach with ASM results in the transverse OAM conservation.

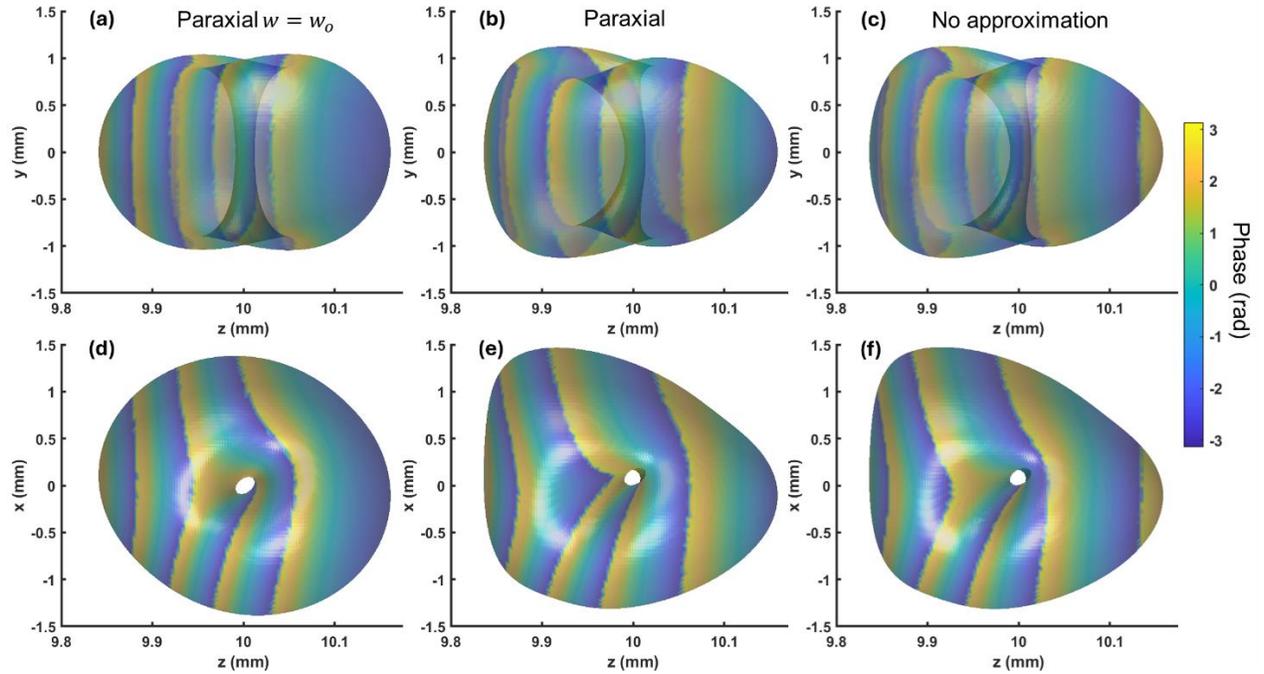

**Figure 3:** STOV intensity profile with phase at 10 mm propagation distance to highlight differences that occur for different approximation approaches. For small bandwidth, the $\omega = \omega_o$ approximation fails to predict the bending singularity which clearly occurs for paraxial (with variable $\omega$) and no approximations.

The intensity profiles of the fields after propagation can appear nearly identical under certain conditions, such as when propagation fields with small spatial and temporal frequency bandwidths. Figure 3 compares the intensity profiles produced from Equation (9) (Figure 3 (a), (d)), Equation (10) (Figure 3 (b), (e)), and 11 (Figure 3 (c), (f)) for parameters $w_k = 1/\text{mm}$, $w_\omega = 1$ THz, $\omega_0 = 4$ THz $A = 1 \text{ mm}^{-1}\text{THz}^{-1}$, $B = 0$ which is a HG like field evolves into an STOV.

Under these conditions, the intensity profile obtained using the small-bandwidth approximation differs noticeably from those produced by the other propagation approaches. The phase singularity line of the STOV bends along with the wavefront in the $yz$ plane (Figure 2 (b) and (c)) and the spiral phase also is distorted by the curving wavefront in the $xz$ plane (Figure 2 (e) and (f)). This distortion is due to the frequency dependent diffraction: as the polychromatic frequency component diffract differently, their combined result bends the phase singularity line, which ensures the conservation of the transverse OAM. Assuming a fixed frequency (small frequency bandwidth approximation) eliminates curvature in space and time, keeping the STOV planar, which can give the misleading impression of a varying OAM. Therefore, for accurate OAM calculations, it is crucial to consider the optical wavepacket in three dimensions with appropriate approximations.

For propagation without approximations using Equation (9), a notable feature is that the slope of the OAM variation decreases as the numerical resolution increases. Figure 4(a) shows the effect of decreasing the z-spacing when computing the gradient in Equation (12) while keeping the $x$-spacing fixed. Figure 4 (b) plots the slopes of the OAM change / time versus resolution for different $x$ resolutions, illustrating that the slope decreases as $z$ and $x$ spacing are reduced. Accurate calculation of $L_y$ therefore requires suitably high resolution in both $z$ and $x$ spacing.

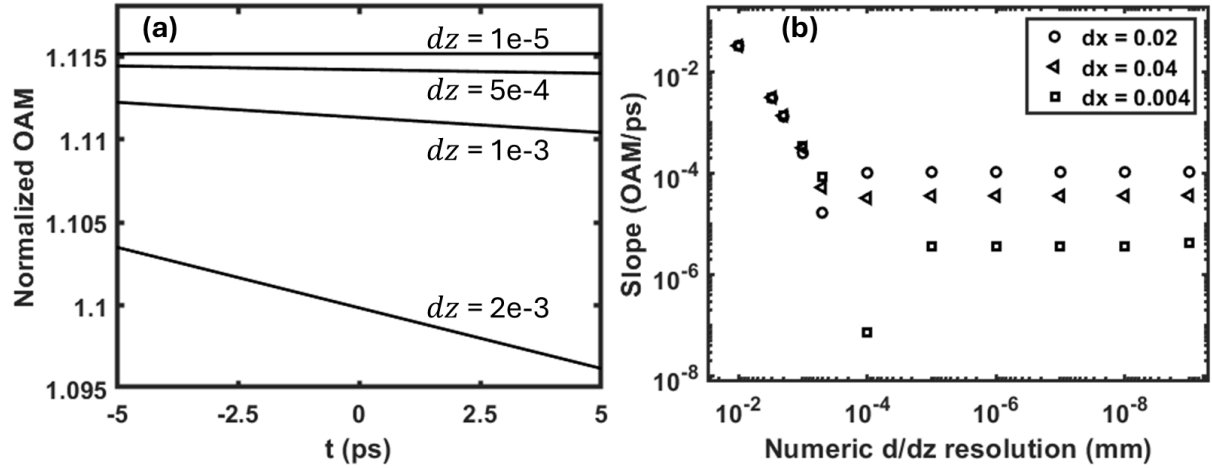

**Figure 4:** The total OAM of the STOV for different resolutions in the numerical derivative for calculating momentum. (units are in mm)

## 3 Conclusion

In conclusion, we have demonstrated that transverse OAM of STOVs is conserved with the standard OAM definition when an appropriate propagation method is used. While paraxial and small bandwidth approximation can provide convenient analytic results, they may produce a spurious variation in OAM. Therefore, it is crucial to consider three-dimensional propagation of the optical wavepacket using appropriate approximations. It appears that the conservation of the OAM is very sensitive to the frequency dependent diffraction. For numerical simulations, high resolution is recommended to accurately observe the OAM conservation. We believe this work will provide valuable insight into the nature of OAM in STOVs.

## 4 Methods

Analytic equations were derived using MATLAB symbolic toolbox. Numeric calculations of the fields and OAM were also performed on MATLAB. The linear momentum was calculated using MATLAB gradient function. For calculating spatial OAM of the STOV, we used a 2001x5x301 grid in $x, y$, and $t$ for calculating the time varying field at each $z$ point. For calculating the spatiotemporal OAM of the STOV, we used 301x5x5001 grid in $x, y$ and $t$. The spatial frequency bandwidth in x and y was 1 mm$^{-1}$ and the center frequency was 4 THz. The spatial span was 5 mm in x and y. The span for z was 2.5 mm and was sampled at 201 points. To calculate the gradient at $z$, two additional points were generated at $z - dz$ and $z + dz$. Figure 2 data was generated with $dz = 1e - 6$ mm.

## Data availability

Data underlying the results presented in this paper are available from the corresponding authors upon reasonable request.

## Author contribution

J. Adams conducted theoretical and numerical simulations. Y. Park contributed for developing theories. A. Chong supervised the project. J. Adams and A. Chong wrote the manuscript.

## Competing interests (mandatory)

Authors declare no competing interests.

## Acknowledgements (optional)

This work is supported the National Research Foundation of Korea (NRF) funded by the Korea government (MSIT) [Grant No. 2022R1A2C1091890] and Global-Learning & Academic research institution for Master's·PhD students, and Postdocs (G-LAMP) Program of the National Research Foundation of Korea(NRF) grant funded by the Ministry of Education [Grant No. RS-2023-00301938] and This study was supported by the Quantum Science Technology Center (QSTC).